\documentclass[prb,twocolumn,aps,showpacs,showkeys,10pt,superscriptaddress,nofootinbib]{revtex4}
\usepackage{natbib}
\usepackage{url}
\usepackage{bm}
\usepackage{amsmath}
\usepackage{ulem}
\usepackage{amsmath}
\usepackage{amssymb}
\usepackage{graphicx}
\usepackage{subfigure}
\usepackage{color}
\usepackage{physics}
\usepackage{braket}
\usepackage{tabularx}
\usepackage[colorlinks=true,linkcolor=red,urlcolor=blue,citecolor=blue]{hyperref}

\begin{document}

\title{Importance of bond exchange in MnC structural stability and
  half-metallic ferromagnetism: a comprehensive benchmark density
  functional study}

\author{Abdesalem Houari}
\email[corresponding author:]{abdeslam.houari@univ-bejaia.dz}
\affiliation{Theoretical Physics Laboratory, 
             Department of Physics, 
             University of Bejaia, 
             Bejaia, Algeria}
\author{Peter E. Bl\"ochl}
\affiliation{Clausthal University of Technology, Institute for
Theoretical Physics, Leibnizstr.10, D-38678 Clausthal-Zellerfeld,
Germany}

\date{\today} 

\begin{abstract}

  	Recently, the first successful synthesis of bulk manganese
        monocarbide (MnC) has been reported. The compound crystallizes
        in the zincblende (B3) structure and has been proposed as a
        potential superhard material. In this work, we present a
        comprehensive first-principles study of the structural,
        electronic, and magnetic properties of MnC using several
        exchange–correlation approximations. We demonstrate that the
        SCAN meta-GGA and the PBE0r hybrid functional outperform
        GGA–PBESol, HSE06, and the mean-field DFT+$U$+$V$ schemes in
        correctly predicting the experimentally observed zincblende
        structure as the ground state. Our analysis reveals that the
        bond-exchange interaction plays a decisive role in stabilizing
        this phase. At theoretical equilibrium, hybrid functionals
        (PBE0r and HSE06) predict a metallic ground state, whereas
        semilocal functionals (PBEsol and SCAN) and DFT+$U$+$V$ yield
        two distinct types of half-metallic ferromagnetism with
        opposite volume dependence, opening a debate on the origin and
        nature of this behavior. If confirmed experimentally, such a
        property would make MnC an attractive candidate for spintronic
        applications. 
\end{abstract}



\maketitle

\section{Introduction}
\label {Intro}


   	The stoichiometric binary mono-compounds arising from late
        3{\it d} transition metal (Mn, Fe, Co, Ni and Cu) and light
        elements (C, N and O) show a variety of interesting physical
        properties.  Among them, the monoxides (e.g. MnO, FeO, NiO),
        are well known since a long time and their properties are well
        established experimentally and theoretically
        ~\cite{boer37,mott37,fujimori84,fujimori88}. These systems,
        which crystallize in the rocksalt structure, have served for
        decades as benchmark materials for the study of strong
        correlations~\cite{sawatzky84,zaanen85}.

   	In contrast, the corresponding mononitrides remained more
        challenging experimentally until the end of the past
        century~\cite{leineweber00,suzuki93,suzuki95,suzuki00}. For
        instance, the experimentally observed antiferromagnetic MnN,
        in a distorted rocksalt structure, is currently well
        established~\cite{houari08,chan08}. On the other side, the
        crystal structure FeN and CoN was the subject of some debates
        ~\cite{houari07,miao07}, but it has been confirmed later that
        both adopt a zincblende one~\cite{soni12}. Finally, NiN and
        CuN remain experimentally unknown, and some {\it ab-initio}
        predictions of their ground state structure continue to
        disagree~\cite{wang04,paduani08,houari19_bis}.

        The situation is more complicated for monocarbides (MnC, FeC
        etc.) which are hardly known. Their difficult synthesis is
        mainly attributed to the progressive filling of the
        antibonding states along the late 3{\it d} series, which
        weakens the metal–carbon bonding and makes the stoichiometric
        phases highly unstable~\cite{singh92,qwang16,haglund93}.
        Nevertheless, some {\it ab initio} investigations have
        addressed their structural and thermodynamic stability in the
        rocksalt phase~\cite{gustev03,szymanski19}. It has been
        concluded that MnC and FeC are unstable at ambient conditions,
        but could be stabilized under high pressure or
        temperature. Interestingly, an unconventional
        superconductivity was also predicted in
        MnC~\cite{szymanski19}. A similar study has demonstrated that
        MnC would be more stable in hexagonal wurtzite
        structure~\cite{khatri21}.

        According to the calculated formation enthalpies of the
        manganese-carbon system, the stoichiometric MnC formation is
        highly unfavorable~\cite{djurovic10}. Recently, however, a
        first successful synthesis was achieved under high-pressure
        and high-temperature conditions~\cite{aparajita18}. The
        compound was found to be a metal adopting the fourfold
        coordinated zincblende (ZB) structure, with a lattice constant
        of 4.429~\AA, and suggested as a potential superhard
        material. However, no data have been provided about its
        magnetic properties. This experimental achievement presents a
        theoretical puzzle, since recent calculations concluded that
        under pressure, MnC would be stable in the rocksalt
        structure~\cite{szymanski19}. Moreover, early theoretical
        investigations predicted zincblende MnC as half-metallic
        ferromagnet with an unusual a coexistence of localized and
        itinerant magnetic
        moments~\cite{sasioglu05,pask03,qian04}. This property
        (e.g. full spin polarization at Fermi level) is
        highly desirable for spintronic
        applications~\cite{houari18,pickett01}.

        This raises a fundamental question about the accurate
        first-principles framework to simultaneously describe the
        structural stability, electronic, and magnetic properties of
        MnC. The answer is non-trivial, as it requires a method
        capable to capture correctly the interplay between covalent
        Mn-C bonding, electronic correlations, and exchange
        interactions.

        Motivated by these developments, we present a comprehensive
        DFT study of MnC. We employ a hierarchy of
        exchange-correlation functionals, ranging from semilocal GGA and
        meta-GGA to range-separated hybrids and the extended
        DFT+$U$+$V$ approach, with a specific focus on the role of
        {\it bond-exchange} interactions. Our goals are threefold: to
        identify the functional that correctly stabilizes the observed
        ZB phase, to uncover the physical mechanism behind this
        stabilization, and to resolve the conflicting predictions
        regarding its electronic behavior.

        The paper is organized as follows: Sec.~\ref{methods} outlines
        the theoretical methods, where the basics of meta-GGA, hybrid
        DFT functionals, and the extended DFT+$U$+$V$ approach are
        presented. The computational details are given in
        Sec.~\ref{comput}. The structural stability and the magnetic
        ordering are discussed in Sec.~\ref{struct}, and we analyze
        the electronic structure and chemical bonding in
        Sec.~\ref{elec}. To conclude our investigation, we summarize
        the important results in Sec.~\ref{conclusion}.

   \section{Theory}
   \label{methods}

   	Although density functional theory (DFT) is, in principle, an
        exact ground-state theory, the exact form of the
        exchange-correlation energy functional $E_{\rm xc}$ remains
        unknown~\cite{hohenberg64}. Therefore, the practical usage of
        the Kohn-Sham framework~\cite{kohn65} relies entirely on
        finding suitable approximations, and a panoply of schemes and
        parameterizations have been developed over time. The
        classification of the main theoretical advances is nicely
        illustrated by the Perdew Jacob's ladder~\cite{perdewjl},
        which ranks functionals by their degree of nonlocality.

   \subsection{meta-GGA}
   \label{mtgga}

   	The so-called meta-generalized gradient approximations
        (meta-GGAs) constitute the highest semilocal rung of Jacob’s
        ladder hierarchy avoiding the full non-locality. Like GGAs,
        they depend on the electron density $n_{\sigma}({\bf r})$ and
        its gradient $\nabla n_{\sigma}({\bf r})$, but they
        additionally incorporate the positive kinetic-energy density
        $\tau_{\sigma}({\bf r})$\footnote{Notice that in addition, the
        Laplaciens of the electron densities $\nabla^2n_{\sigma}({\bf
          r})$, can be incorporated into the meta-GGA functionals
        \cite{perdew99}}, where $\sigma =\{\uparrow, \downarrow \}$ is
        the spin index~\cite{perdewjl}: 

   \begin{equation}\label{xc_mtgga}
     E_{\rm xc}^{\rm MTGGA} \left[n_{\sigma} \right] =
     \int d{\bf r} \varepsilon_{xc}^{\rm MTGGA} (n_{\sigma},\nabla n_{\sigma},\tau_{\sigma})
   \end{equation}

   	The kinetic-energy densities are constructed from the
        Kohn-Sham orbitals :

   \begin{equation}\label{kin_mtgga}
     \tau_{\sigma}({\bf r}) = \sum_{i}^{\rm occ}\frac{1}{2}
     |\nabla \phi_{i}^{\sigma} ({\bf r})|^2
   \end{equation}

   	Several parameterizations of the meta-GGA exist in literature,
        where the well known are the TPSS~\cite{tao03}, its revised
        version rev-TPPS~\cite{perdew09}, M06L~\cite{zhao06} and
        mBJ~\cite{tran09}.  The recently proposed one, namely the
        strongly constrained and appropriately normed (SCAN)
        functional~\cite{sun15}, has attracted much
        attention. According to the authors, the most important
        ingredient making the SCAN superior to previous meta-GGAs, is
        the dimensionless parameter of the kinetic-energy densities:

   \begin{equation}\label{kin_tau}
     \alpha = \frac{\tau_{\sigma}({\bf r}) - \tau_{\sigma}^{\rm W}({\bf r})}{\tau_{\sigma}^{\rm unif}({\bf r})} > 0 ,
   \end{equation}

   	This parameter distinguishes three different bonding
        characters: covalent ($\alpha = 0$), metallic ($\alpha =1$)
        and weak ($\alpha \gg 1$). Many SCAN calculations, on diverse
        molecular and extended systems, have demonstrated that it
        leads to very interesting results; especially for structural
        geometry and
        energetics~\cite{sun16,zhang18,isaacs18,tran16,sokolov23}.
        Moreover, it is argued that SCAN functional could be suitable
        (at least qualitatively) for the treatment of strong
        electronic correlations~\cite{lane18,fu19}, and the related
        ground-state symmetry breaking~\cite{perdew21}. We notice
        finally that two revised formulation of the SCAN functional,
        known as rSCAN~\cite{bartok19} and r2SCAN~\cite{furness20},
        have been recently proposed. 
   
   \subsection{Hybrid (Hartree-Fock)-DFT}
   \label{hybrid}

   	An outstanding improvement, beyond semilocal DFT, is the {\it
          hybrid} functionals. The basic idea, initially proposed by
        Becke \cite{becke1,becke2}, is to incorporate a fraction of
        nonlocal exact exchange from Hartree–Fock (HF) theory into the
        semi-local $E_{\rm xc}$. He suggested a simple linear
        combination:

   \begin{equation}\label{becke}
    E_{\rm xc}^{\rm Hybrid} = E_{\rm xc}^{\rm GGA} + \alpha \left(E_{\rm x}^{\rm HF} - E_{\rm x}^{\rm GGA} \right)
   \end{equation}

   	where $\alpha$ is obtained by some fitting to the atomization
        energies. A later improved version, the B3LYP
        functional\cite{B3LYP}, has become very popular in quantum
        chemistry community. A next step is the PBE0, described as a
        parameter-free functional~\cite{PBE0}. Based on
        GGA-PBE\cite{perdew96}, it is derived from equation
        \ref{becke} with a predefined fraction ($\alpha =
        \frac{1}{4}$), justified by a general analysis from
        perturbation theory \cite{perdewPBE0}:
   
   \begin{equation}\label{PBE0}
    E_{\rm xc}^{\rm PBE0} = E_{\rm xc}^{\rm PBE} + \frac{1}{4} \left(E_{\rm x}^{\rm HF} - E_{\rm x}^{\rm PBE} \right)
   \end{equation}

   	While the previous functionals belong to the family of {\it
          global} hybrid functionals, an important advance is the
        so-called {\it range-separated} hybrid functionals, using a
        {\it screened} interaction instead of the bare Coulomb
        potential~\cite{HSE03,HSE06}. In extended systems, this idea
        is very appealing not only technically but also physically.  A
        {\it screened} interaction makes the HF exchange more
        tractable, and accounts for the screening phenomenon
        established by the random-phase
        approximation~\cite{RPA1,RPA2}.

        In HSE hybrid functionals~\cite{HSE03,HSE06}, the separation
        (and the screening as a result) is achieved by an error
        function decomposing the bare Coulomb potential into a
        short-range (SR) and a long-range (LR) parts:

   \begin{equation}\label{erf}
    \frac{1}{r} = \underbrace{\frac{\rm 1 - erf( \omega r )}{r}}_{\rm SR}  + \underbrace{\frac{\rm erf( \omega r )}{r}}_{\rm LR}
   \end{equation}

   	where $\omega$ is an adjustable screening
        parameter~\cite{krukau}. The HSE exchange-correlation
        functional is expressed as: 

   \begin{equation}\label{HSE}
     E_{\rm xc}^{\rm HSE} = \frac{1}{4} E_{\rm x}^{\rm HF-SR} + \frac{3}{4}E_{\rm x}^{\rm PBE-SR}
                         + E_{\rm x}^{\rm PBE-LR}  + E_c^{\rm PBE}
   \end{equation}

   	In fact, GGA-PBE and PBE0 hybrid functional are merely
        restored, from the two particular values $\omega = \infty$ and
        $\omega = 0$, respectively.  In equation~\ref{HSE}, only the
        first term needs to be calculated exactly, and all the
        remaining ones are computed at the PBE level. Even with
        this simplification, the HF exact
        exchange is computationally very expensive. With the Kohn-Sham
        orbitals $\phi_{n,{\bf k}}({\bf r})$, it is defined as:

   \begin{equation}\label{HF-SR}
     \begin{aligned}
     E_{\rm x}^{\rm HF-SR} =\displaystyle & -\frac{1}{2} \sum_{\sigma} \sum_{n,{\bf k}}\sum_{n^{\prime},{\bf k^{\prime}}}
     \int\int d{\bf r} d{\bf r^{\prime}} \left(1 - \rm erf(\omega |{\bf r} - {\bf r^{\prime}}|) \right) \\
      &\times \frac{\phi_{n,{\bf k}}^{\sigma *}({\bf r})\phi_{n^{\prime},{\bf k^{\prime}}}^{\sigma}({\bf r})
                 \phi_{n^{\prime},{\bf k^{\prime}}}^{\sigma *}({\bf r^{\prime}})\phi_{n,{\bf k}}^{\sigma}({\bf r^{\prime}})}
                {|{\bf r} - {\bf r^{\prime}}|}
     \end{aligned}
   \end{equation}

   	Another progress in this field is the recently invented PBE0r
        hybrid functional\cite{blochl11}, following a new and different
        approach. The Kohn–Sham orbitals are
        expanded in a minimal basis of localized atom-centered
        tight-binding orbitals, allowing a natural decomposition of
        the HF exchange into:
   
   	\begin{itemize}
   	\item {\it On-site} exchange which acting between states
          centered on the same atom. It affects the degenerate
          states and split them into filled and empty bands (the
          Mott-Hubbard separation in the narrow and
          partially filled {\it d} and {\it f} atomic shells).
          \item {\it Bond-exchange} describing the
            intersite exchange between electron densities
            located at neighboring atoms involved in chemical
            bonds.  
         \item {\it Long-range} exchange acting over distances longer
           than bond lengths. The neglect of this term is indeed
           advantageous, since the unphysical behavior in metallic
           solids (due to the long-range tail of the Coulomb
           interaction) is avioded and the screening is effectively
           taken into account.
        \end{itemize}

        The {\it bond exchange} term is particularly relevant for
        covalent materials, as it differentiates between bonding and
        antibonding states. It mainly corrects the semilocal
        functionals errors in formation-energy and band gaps.

        The use of localized atom-centered orbitals (a natural cutoff
        of the exchange interaction) and the neglect of the long-range
        terms (capturing the important screening effects) make PBE0r a
        {\it de facto} range-separated functional.  Its {\it local}
        (i.e. on-site) formualtion is defined as:
   
     \begin{equation}\label{PBE0r}
      E_{\rm xc}^{\rm PBE0r} = E_{\rm xc}^{\rm PBE} + \sum_{n=1}^N a_n
      \left(E_{x,n}^{\rm HF-onsite} - E_{x,n}^{\rm PBE-onsite} \right) 
     \end{equation}

     	The sum runs over the $N$ atoms of the system (i.e. the cell
        used in calculations), where $a_n$ represents the mixing
        factor (fraction) of the HF exact exchange. This formulation
        is very flexibile, since the mixing factor $a_n$ can be set
        differently for the atomic species; and allows an
        explicit control of the bond-exchange contribution.

        Finally, the accuracy of PBE0r versus experiments has been
        established for a large variety of systems; such as partially
        filled {\it d}-shell oxides~\cite{sotoudeh17},
        transition metal sulfides and nitrides
        ~\cite{houari18,houari19,benissad20},
        manganese oxide~\cite{eckhoff20,allaoua25}, and
        iron spin-crossover
        complexes~\cite{rajpurohit24}.
     


   \subsection{Hubbard corrected DFT}
   \label{hubbard}

   	In this section, we summarize the basic notions of the
        Hubbard-corrected DFT methods. We concentrate on the recently
        proposed DFT+$U$+$V$ formalism\cite{campo10} (based on the
        extended Hubbard model), from which the widely used 
        DFT+$U$~\cite{anisimov91,liechtenstein95,dudarev98} can be
        straightforwardly derived.  Starting from any local or
        semi-local functional, the corresponding DFT+$U$+$V$ one is
        written as:

   \begin{equation}\label{DFT+U+V}
      E_{\rm xc}^{{\rm DFT}  +U + V} = E_{\rm xc}^{\rm DFT} + E_{\rm xc}^{U + V} 
   \end{equation}

   	Following the definition and the notation adopted in
        references~\cite{campo10,timrov22_prx}, the Hubbard correction
        to the exchange-correlation energy (in the simplified
        rotationally-invariant formulation) is given by:

   \begin{equation}\label{E_U+V}
     \begin{aligned}
     E_{\rm xc}^{U+V} &= \displaystyle  \frac{1}{2} \sum_I \sum_{\sigma m m^{\prime}}
     U^I \left(\delta_{m m^{\prime}} - n_{m m^{\prime}}^{II\sigma} \right)n_{m^{\prime} m}^{II\sigma} \\
     & - \frac{1}{2} \sum_I \sum_{J\neq I}^{n}\sum_{\sigma m m^{\prime}} V^{IJ}n_{m m^{\prime}}^{IJ\sigma}n_{m^{\prime} m}^{JI\sigma}
     \end{aligned}
   \end{equation}

   	where $U^I$ and $V^{IJ}$ are, respectively, the on-site and
        the intersite Hubbard parameters; and the indices $I$ and $J$
        run over all the atomic sites. Notice that the sum over $J$ in
        the second line, covers the all neighboring atoms included in
        the intersite correction.

        In equation~\ref{E_U+V}, the onsite occupation matrix elements
        $n_{mm^{\prime}}^{II\sigma}$ are obtained by the projection of
        the Kohn-Sham states $\ket{\psi_{{\bf k}\nu}^{\sigma}}$ on a
        set of localized (atomic) orbitals $\ket{\phi_{m}^{I}}$. 

   \begin{equation}\label{occ_mat}
     n_{mm^{\prime}}^{IJ\sigma} = \sum_{{\bf k}\nu} f_{{\bf k}\nu}^{\sigma} \braket{\psi_{{\bf k}\nu}^{\sigma}|\phi_{m^{\prime}}^{J}}
     \braket{\phi_{m}^{I}|\psi_{{\bf k}\nu}^{\sigma}}
   \end{equation}

   	where $f_{{\bf k}\nu}^{\sigma}$ represent the occupations of
        the Kohn-Sham states, $\nu$ is the electronic band index and
        ${\bf k}$ is the Bloch wave vector. The $m$ ($m^{\prime}$)
        index is the magnetic quantum number, running over the orbital
        momentum in atom $I$ ($J$). Notice that the on-site occupation
        elements $n_{mm^{\prime}}^{II\sigma}$ can be obtained as a
        particular case ($I=J$).

        At this point, it is important to mention the issue about the
        values of $U$ and $V$ parameters. These latter are actually
        unknown, and they are usually tuned by hand to reproduce some
        experimental data (like lattice constants, magnetic moments
        and more importantly band gaps). Such a procedure, however,
        makes DFT+ $U$ + $V$ a partially empirical method. In fact,
        there exist several first-principles techniques to
        compute the correct values of $U$ and $V$. Some of the well
        known are the constrained DFT
        (cDFT)~\cite{gunnarsson89,hybertsen89,gunnarsson90} and the
        constrained random-phase approximation
        (cRPA)~\cite{springer98,kotani00,aryasetiawan04}. A recent
        one, using density-functional perturbation theory
        (DFPT)~\cite{timrov18,timrov21}, has been developed with the
        main advantage of a significant reduction in the computational
        cost.

        Last but not least, these Hubbard parameters depend intimately
        on the chemical environment from which they are
        computed. Thus, they are not transferables i.e. not universal
        for a given atomic species~\cite{kulik08}. In addition, they
        depend on the kind of the used projector
        functions~\cite{lowdin50,wang16}.

   \section{Computational details}
   \label{comput}

   	   All calculations were performed within
           DFT~\cite{hohenberg64,kohn65}, using three different
           exchange-correlation schemes. The semi-local framework
           (GGA-PBESol~\cite{perdew08} and SCAN
           meta-GGA~\cite{sun15}), Two range-separated hybrid
           functionals (HSE06~\cite{HSE06} and PBE0r ~\cite{blochl11})
           and the mean-field DFT+$U$+$V$ approach~\cite{campo10}.  In
           the latter, the double counting correction is accounted
           using the {\it fully localized limit}
           formulation~\cite{anisimov93,czyzyk94,solovyev94,liechtenstein95}.

           The PBESol, SCAN, HSE06 and DFT+$U$+$V$ calculations were
           carried out using the Quantum ESPRESSO package (QE, version
           7.1)~\cite{giannozzi09,giannozzi17,giannozzi20}. In the
           PBESol, HSE06 and DFT+$U$+$V$, the Kohn-Sham
           single-particle equations were solved within the
           projector-augmented wave (PAW) method~\cite{paw}, using PAW
           potentials from the PSlibrary \cite{pslibrary}. For SCAN
           meta-GGA, however, norm-conserving pseudopotentials from
           the PseudoDojo library~\cite{psdojo} were employed.

           Based on density-functional perturbation theory (DFPT),
           within the linear-response formalism; the on-site $U$ and
           intersite $V$ parameters were calculated self-consistently
           using the HP code~\cite{timrov22} (distributed within the
           QE package). The $\Gamma$-centered {\bf q}-point mesh was
           increased up to $\rm 3 \times 3 \times 3$ to ensure
           convergence. Orthogonalized atomic orbitals, obtained via
           L\"owdin method~\cite{lowdin50,timrov22,mayer02}, were used
           as Hubbard projectors functions. Since $U$ and $V$ are not
           transferables, they were computed separately for the
           rocksalt and zincblende structures, at each
           step of the total-energy versus volume
           optimization.

           All calculations were converged with a plane-wave
           kinetic-energy cutoff of~50 Ry, with a total-energy
           threshold of $10^{-8}$ Ry in the self-consistent cycle.
           The Brillouin zone was sampled following the Monkhorst-Pack
           scheme~\cite{monkhorst76}, using a $18 \times 18 \times
           18 $ mesh (200 {\bf k}-point).
           For metallic states, the electronic smearing was treated
           using the Marzari-Vanderbilt scheme~\cite{marzari99}, with
           a broadening parameter of 0.02~Ry. The calculations of the
           electronic densities of states, however, were carried out
           with the improved linear tetrahedron
           method~\cite{bloechl94}.

           The PBE0r hybrid functional calculations were performed
           using the PAW method \cite{paw}, within its original
           implementation provided in the {\tt CP-PAW}
           code~\cite{cp-paw}. We recall that PAW is by construction
           an {\it all-electron} method, using augmented-waves.  The
           included orbitals in the PAW augmentation are: 2{\it s},
           2{\it p}, and 3{\it d} of C, and 3{\it s}, 3{\it p}, 3{\it
             d}, 4{\it s}, and 4{\it p} of Mn; with covalent radii of
           0.71 \AA~ (C) and 1.17 \AA~ (Mn). In addition, Mn 3{\it s}
           and 3{\it p} semi-core states were treated as valence
           orbitals. As mentioned above in section \ref{hybrid}, PBE0r
           uses a localized tight-binding orbitals basis, including
           the 2{\it s} and 2{\it p} orbitals of C, and the 3{\it s},
           3{\it p}, 3{\it d}, and 4{\it s} ones of Mn.

           The {\tt CP-PAW} code uses the Car-Parrinello {\it ab
             initio} molecular dynamics (AIMD)\cite{car}, where the
           ground-state is reached by applying a friction term (a
           damped MD). Here also, and to be in line with QE
           calculations, a plane-wave kinetic-energy cutoff of~50 Ry
           for the auxiliary wave functions was used. The total energy
           was minimized up to a convergence threshold of 10$^{-5}$
           Hartree, and the improved linear tetrahedron method
           \cite{bloechl94} was used for the Brillouin zone
           integration (with 200 {\bf k}-points). To deal with fractional
           occupations in metallic systems, the Mermin
           functional~\cite{mermin} was used.

           A significant advantage of PBE0r functional is its low
           computational cost, that is comparable to
           GGA~\cite{eckhoff20,allaoua25}. In the present study, the
           same value of the mixing factor is assigned for both C and
           Mn atoms. As mentioned above (section~\ref{hybrid}),
           however, HSE06 calculations are extremely expensive;
           especially with plane-waves~\cite{timrov22_prx}. To assess
           the exact exchange effect on MnC properties, we have used
           two fractions of the exact exchange: 10\% and 30\% , with a
           $\rm 3 \times 3 \times 3$ uniform {\bf q}-point mesh. 


   \section{Ground-state crystal structure and magnetic order}
   \label{struct}
   %
   %
   \begin{figure*}[htb]
     \includegraphics[width=0.78\textwidth,height=11.5cm]{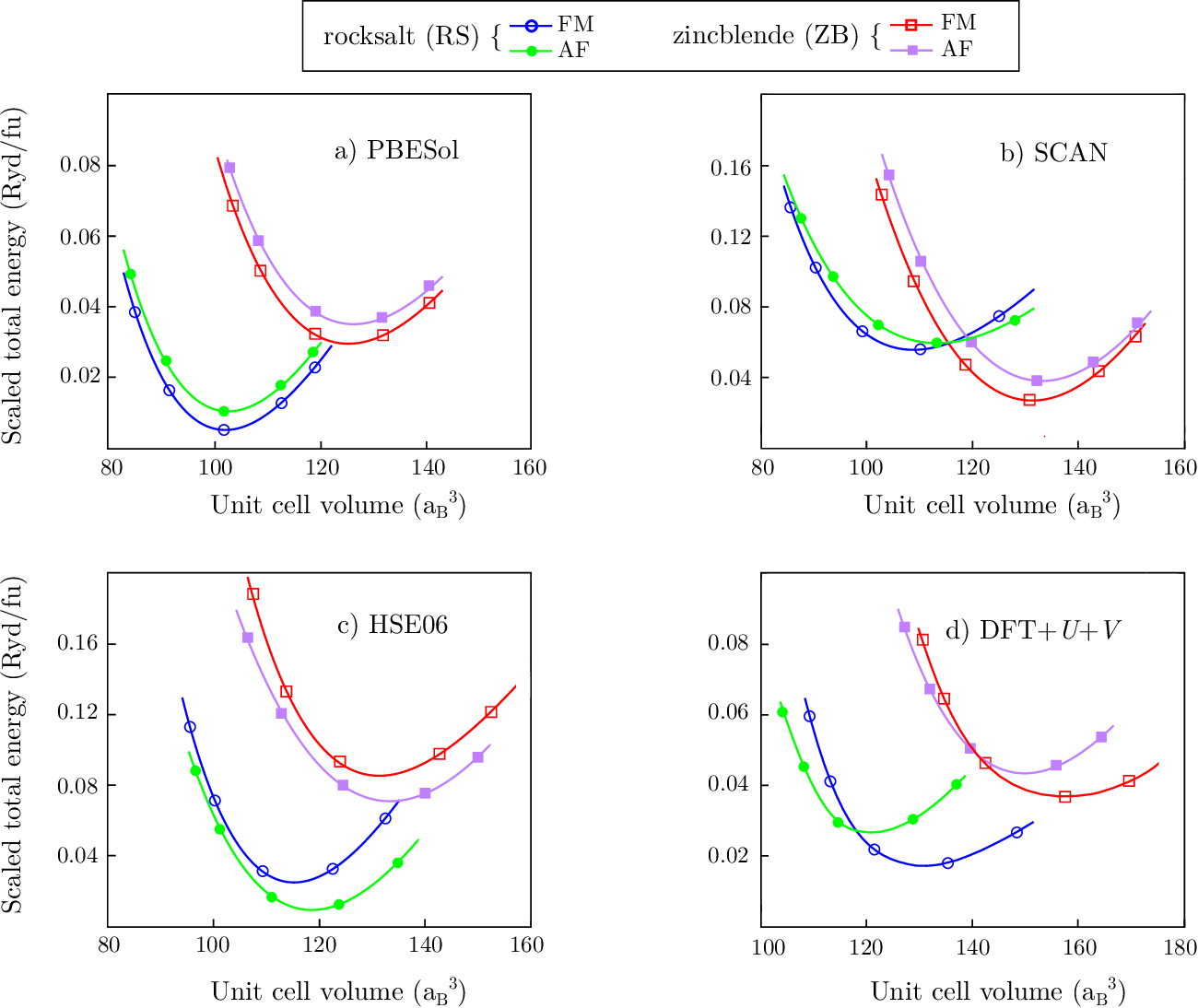}
     \caption{Energy versus volume optimization of MnC in rocksalt (RS)
       and zincblende (ZB) structures; as obtained within QE:
       a) PBESol, b) SCAN, c) HSE06 and d) DFT+$U$+$V$ functionals.}
     \label{fig1:blk_all}
   \end{figure*}
   \begin{figure*}[htb]
     \includegraphics[width=0.80\textwidth,height=6.2cm]{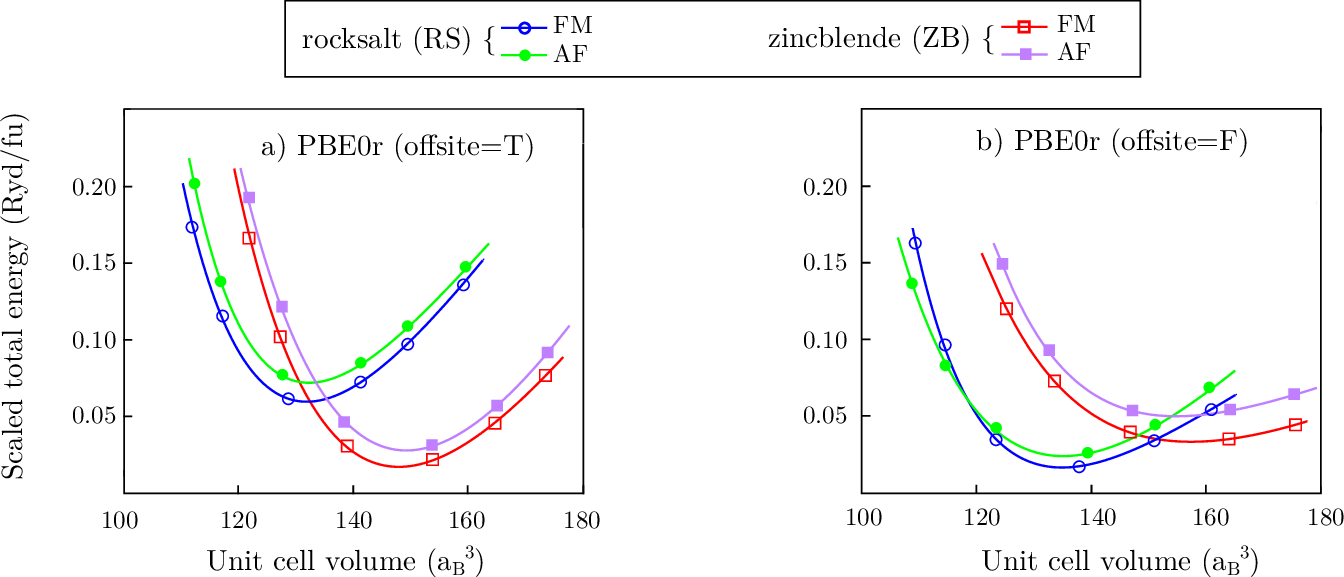}
     \caption{Energy versus volume optimization of MnC in rocksalt (RS)
       and zincblende (ZB) structures; as obtained within
       PBE0r functional: a) PBE0r (offsite=T) with the bond exchange, and b)
       PBE0r (offsite=F) without the bond exchange.}
     \label{fig2:blk_pbe0r_on-vs-off}
   \end{figure*}

   	Structural stability of MnC is investigated in the two
        faces-centered cubic: rocksalt (RS) and zincblende (ZB)
        structures. This choice is motivated by the fact that all
        known monoxides, mononitrides and monocarbides are found to
        crystallize in either of these two prototypes. For each
        structure, both spin-polarized ferromagnetic (FM) and type-II
        antiferromagnetic (AF-II) are considered. In the latter,
        ferromagnetic planes alternate along the $\langle \rm 1 1 1
        \rangle$ direction, as commonly observed in MnO and NiO.
 
   	As an aside note, additional calculations for MnC in the
        wurtzite structure have been performed. It was found to be the
        most unstable phase and the corresponding results are
        therefore omitted.

   \subsection{Structural stability}
   \label{stability}
   
   %
   %
   \begin{table}[htb]
     \caption{Calculated total energies differences (in Ryd/fu)
       between FM and AF-II orders in the rocksalt (RS) and zincblende
       (ZB) structures, and between the two phases at theoretical
       equilibrium; as arising from all the frameworks.}
     \begin{ruledtabular}
       \begin{tabular}{l|cc|c}
         
                   &$\rm E_{\rm FM} - E_{\rm AF-II}$ &   &$\rm E_{\rm RS} - E_{\rm ZB}$   \\
         \cline{2-3}
                   & RS          &        ZB      &    \\ 
   \hline
   PBESol          & $-$ 0.003   &   $-$ 0.005   &  $-$ 0.028   \\
   
   SCAN            & $-$ 0.006   &   $-$ 0.008   &   + 0.035   \\
  
   PBE0r           & $-$ 0.015   &   $-$ 0.010   &   + 0.043   \\
  
   HSE06           & + 0.020     &   + 0.015      &   $-$ 0.045  \\

   DFT+$U$+$V$     & $-$ 0.008   &   $-$ 0.005    &   $-$ 0.029  
       \end{tabular}
       \label{table1}
     \end{ruledtabular}
   \end{table}
   \begin{table*}[htb]
     \caption{\label{table2} Calculated equilibrium properties within
       all the used frameworks, for $ {\rm {MnC}}$ in rocksalt (RS)
      and zincblende (ZB) structures. The lattice constant $a$ in \AA,
      the magnetic moment per formula unit M$_{\rm fu}$ in $\mu_{\rm
        B}$ (in the FM case) and the bulk modulus B in GPa are shown.}
    \begin{ruledtabular}
      \begin{tabular}{l|lll|lll}
                      &      & RS     &            &       & ZB     &            \\
  \cline{2-7}
                      & $a$  & B     & M$_{\rm fu}$ & $a$   &  B     & M$_{\rm fu}$ \\
  \hline
   \underline{PBESol} &      &     &       &      &     &                        \\
   FM                 & 3.97 & 230 & 1.15  & 4.22 & 172 & 1.00                   \\
   AF                 & 3.98 & 232 &       & 4.23 & 166 &                         \\
   \hline
   \underline{SCAN}   &      &     &       &      &     &                         \\
   FM                 & 4.02 & 224 & 1.30  & 4.28 & 175 & 1.00                    \\
   AF                 & 4.08 & 219 &       & 4.31 & 169 &                         \\
   \hline
  \underline{PBE0r}   &      &     &       &      &     &                         \\
   FM                 & 4.15 & 214 & 2.88  & 4.44 & 192 & 2.05                    \\
   AF                 & 4.17 & 210 &       & 4.48 & 150 &                          \\
  \hline
   \underline{HSE06}  &      &     &       &      &     &                         \\
   FM                 & 4.10 & 238 & 2.58  & 4.32 & 201 & 1.94                    \\
   AF                 & 4.14 & 220 &       & 4.35 & 189 &                        \\
   \hline
   \underline{DFT+$U$+$V$} &      &     &      &      &     &                    \\
   FM                         & 4.27 & 211 & 3.04 & 4.51 & 116 & 3.00            \\
   AF                         & 4.17 & 204 &      & 4.46 & 102 &                  \\
   \hline
   \hline
  \underline{Theory}       &      &     &     &      &     &                \\
  Ref.\cite{khatri21}      & 4.02 & 295 &     & 4.31 & 240 &                  \\
  Ref.\cite{sasioglu05}    &      &     &     & 4.23 &     &               \\
  Ref.\cite{qian04}        &      &     &     & 4.20 &     &                    \\
  \hline
  \underline{Experimental} &      &     &     &      &     &                     \\
  Ref.\cite{aparajita18}   &      &     &     & 4.43 & 170 &                 \\
  
      \end{tabular}
    \end{ruledtabular}
   \end{table*}

   	Theoretical ground-state is obtained from the energy versus
        volume optimization. The results from PBESol, SCAN, HSE06 and
        DFT+$U$+$V$ are shown in Fig.~\ref{fig1:blk_all}, and those
        from PBE0r (with and without the inclusion of {\it bond
          exchange}) are presented in
        Fig.~\ref{fig2:blk_pbe0r_on-vs-off}. The curves were fitted
        with the third-order of Birch-Murnaghan equation of state
        \cite{birch47}. The important finding is that SCAN and PBE0r
        successfully reproduce the experimentally observed zincblende
        ground state structure, while the others incorrectly stabilize
        the rocksalt one. The computed total-energy differences at
        equilibrium are provided in Tab~\ref{table1}. It is important
        to mention that for PBE0r, only a 10\% fraction of exact
        exchange was sufficient to reproduce the correct structure,
        whereas with a fraction as high as 30\% HSE06 still predict a
        stable rocksalt phase.

   	The failure of the GGA-type functionals is actually not new,
        and has been reported in several other cases, such as MnN
        compound~\cite{suzuki00,miao07,hong05}. In that case, however,
        hybrid functionals and GGA+$U$ approaches were able to remedy
        this issue~\cite{chan08}. Interestingly, neither the usually
        accurate HSE06 nor the DFT+$U$+$V$ method are able to predict
        the correct MnC structure.

   \subsection{Role of bond exchange}
   \label{bond}
   
   	This result can be addressed using the flexibility of PBE0r
        framework, where two kinds calculations are shown in
        Fig.~\ref{fig2:blk_pbe0r_on-vs-off}:
        \begin{itemize}
          \item a) with {\it bond}-exchange included (offsite=T,
             that refers to the
            exchange interaction offsite terms)
          \item b) with only on-site exchange (offsite=F)
        \end{itemize}
        
        The PBE0r results provide explicit evidence of the importance
        of the {\it bond}-exchange term in stabilizing the ZB phase.
        When it is ignored, the rocksalt phase becomes energetically
        favored. This exchange term has the characteristic to
        distinguish between bonding and antibonding states, and acts
        differently on their occupations. As a result, it accounts
        better for the covalency effects and reduce the formation
        energy errors.  To assess quantitatively this contribution in
        each phase, the total energy difference $\rm \Delta E =
        E_{(offsite=T}) - E_{(offsite=F)}$ is provided: $\rm \Delta
        E_{RS} = -0.196$ Ryd and $\rm \Delta E_{ZB} = -0.311$
        Ryd. This shows clearly that neglecting the {\it
          bond}-exchange contribution introduces a much larger error
        in the zincblende structure than in the rocksalt one.

        Concerning SCAN functional, its main advantage is the ability
        to recognize different types of bonding. From a
        benchmark calculations (up to 1000 compounds known
        experimentally), it was found that the formation energy errors
        in covalent systems decrease substantially
        (by $\sim$ 50\% compared to GGA-PBE). The reason
        behind this success is the enhanced exchange interaction in
        the covalent regime~\cite{tran09,isaacs18}.
        This agreement between SCAN and PBE0r indicates that the
        enhanced exchange interaction built into SCAN plays a 
        role analogous to the explicit {\it bond}-exchange in PBE0r.

        Both frameworks highlight the important role of covalent
        exchange effects in MnC. In fact, the Mn-C bond involves
        significant hybridization between Mn-3{\it d} and C-2{\it p}
        orbitals; thus, conventional GGAs treating exchange
        isotropically underestimate the energy gain associated with
        covalent bonding.  The explicit inclusion of this term
        (PBE0r) or its implicit treatment via the kinetic-energy
        density parameter (SCAN) both
        strengthen Mn–C hybridization. These findings emphasize that
        accurate treatment of bond-directional exchange is crucial for
        describing structural stability in late 3{\it d}
        transition-metal carbides, where the interplay between
        metallic and covalent bonding is not clear cut.

        We believe that the inability of HSE06 and DFT+$U$+$V$ to
        capture this stabilization can be attributed to the absence of
        such bond-selective exchange. HSE06 mixes a whole and fixed
        fraction of exact exchange into the GGA functional without
        differentiating bonding and antibonding contributions. The
        same argument hold as well for the DFT+$U$+$V$ method.
        Despite the existence of $V$ intersite parameter; it cannot
        fully capture the Mn–C covalency.

   \subsection{Equilibrium Properties and Magnetic Ordering}
   \label{equil}

   	The equilibrium values of lattice constants (in \AA),
        bulk moduli (in GPa) and magnetic moments per formula unit
        M$_{\rm fu}$ (in $\mu_{\rm B}$) obtained from each framework
        are summarized in Tab.~\ref{table2}; along with available
        theoretical and experimental data.

        Compared to the only available experimental lattice constant
        of zincblende MnC~\cite{aparajita18}, semilocal PBESol and
        SCAN functionals tend to underestimate it, while DFT+$U$+$V$
        overestimates it. On the other side, the hybrid functionals
        yield a close agreement with experiment. In particular the
        PBE0r result of 4.44 \AA~ is almost perfect. To our knowledgr,
        MnC has not been elaborated in the rocksalt structure, and our
        PBESol results agree very well with previous GGA
        calculations~\cite{szymanski19,khatri21}.  The bulk moduli
        follow an opposite trend: semilocal functionals agree better
        with experimental results, whereas hybrid and DFT+$U$+$V$
        schemes yield stiffer or softer lattices, respectively.

        Total energies differences between the FM and AF-II states, in
        both crysral structures are provided in Tab.~\ref{table1}.
        The FM order is energetically favored over the AF-II
        configuration in all frameworks, except in HSE06 where the
        opposite is predicted. The computed magnetic moments per formula
        unit reveal clear functional-dependent trends, with larger
        values in the RS phase.  While PBEsol and SCAN yield smaller
        moment, hybrid functionals (HSE06 and PBE0r) enhance the
        exchange splitting and produce larger values. Finally, the
        DFT+$U$+$V$ method yields the largest moments, reflecting a
        stronger on-site exchange interactions.

        The integer magnetic moments obtained within semilocal and
        DFT+$U$+$V$ frameworks suggest a half-metallic nature for
        ZB–MnC. This prediction is examined in more detail in the
        following section.

   \section{Electronic structure and bonding properties}
   \label{elec}

   	In this section, we analyze the electronic structure and
        chemical bonding properties of MnC in both rocksalt (RS) and
        zincblende (ZB) phases. The results are presented in terms of
        the site-, orbital-, and spin-projected densities of states
        (PDOS), as computed within the various exchange–correlation
        frameworks discussed above.  To elucidate the crystal field
        effects, the Mn 3{\it d}-states are splitted into Mn-{\it
          t$_{2g}$} and Mn-{\it e$_g$} manifolds, corresponding to
        octahedral and tetrahedral coordination, respectively.  In
        addition, we concentrate on the ground state spin-polarized
        ferromagnetic configuration; where a particular attention is
        paid to the nature of half-metallic ferromagnetism and to how
        the choice of functional affects the Mn–C bonding and spin
        polarization.

   \subsection{Semi-local PBESol and SCAN}
   \label{pbesol-scan_results}

\begin{figure}[ht]
     \includegraphics[width=0.85\columnwidth,height=12cm]{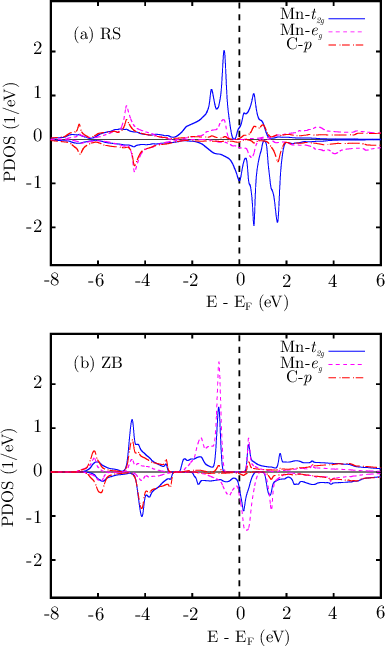}
     \caption{Atom- and spin-resolved densities of states (PDOS) of $
       {\rm MnC} $ in RS-type (a) and ZB-type (b) structures, as
       arising from the PBESol spin-polarized FM calculations.}
     \label{fig3:dos_pbesol}
   \end{figure}

   	First, the MnC electronic structure is explored within PBESol
        functional, where the obtained densities of states are shown
        in Fig.~\ref{fig3:dos_pbesol}. In general, there are three
        main regions corresponding to bonding, non-bonding and
        anti-bonding states.

   	For the rocksalt structure (the upper panel of
        Fig.~\ref{fig3:dos_pbesol}), the bonding region
        [-8 eV, -3 eV] is formed by Mn-{\it e$_g$} and C-{\it
          p} hybridized states with a strong $\sigma$-type overlap
        characteristic of the octahedral arrangement. Next, the
        non-bonding part dominated by the Mn-{\it t$_{2g}$} manifold
        from -3 eV to +1 eV. In the remaining higher energy window,
        the Mn-{\it e$_g$} and C-{\it p} hybridization forms the
        anti-bonding counter-part. The RS phase has clearly a metallic
        behavior, and the small exchange splitting indicates a
        low-spin configuration; with a magnetic moment value of $\sim$1.15
        $\mu_{\rm B}$ arising primarily from the Mn-{\it t$_{2g}$}
        states.

        In the zincblende structure, the crystal-field splitting is
        obviously inverted due to the tetrahedral environment, such
        that the Mn-{\it t$_{2g}$} orbitals form the bonding
        manifold. As displayed in Fig.~\ref{fig3:dos_pbesol} (the
        lower panel), the bonding region [-6 eV, -3 eV] contains
        hybridized Mn-{\it t$_{2g}$} and C-{\it p} states. The
        non-bonding region is dominated by two Mn-{\it e$_g$} bands
        (one occupied and one empty),
        still with a small contribution of Mn-{\it t$_{2g}$}. Above
        +1.5 eV, the anti-bonding region contains essentially C-{\it
          p} and Mn-{\it t$_{2g}$} hybrid states. The large separation
        between bonding and anti-bonding parts of the Mn-{\it
          t$_{2g}$} manifold indicates the high strength of the Mn-C
        bond. More importantly, zincblende MnC is insulating in the
        majority-spin channel with an energy gap of about 1.2 eV,
        while being metallic in the minority-spin one. This yields a
        half-metallic ferromagnet with 1 $\mu_{\rm B}$ magnetic moment
        (from the occupied Mn-{\it e$_g$} band), consistent with previous
        predictions~\cite{pask03,qian04,sasioglu05}.

\begin{figure}[htb]
     \includegraphics[width=0.85\columnwidth,height=12cm]{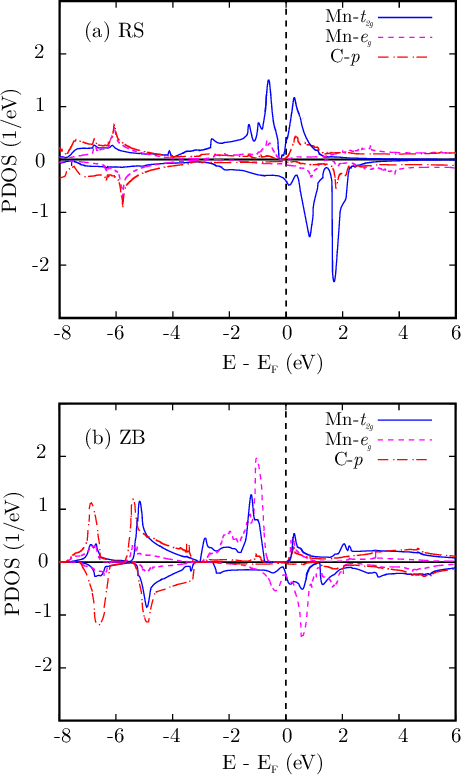}
     \caption{Atom- and spin resolved densities of states (PDOS) of $
       {\rm MnC} $ in RS-type (a) and ZB-type (b) structures, as
       arising from the SCAN spin-polarized FM calculations.}
     \label{fig4:dos_scan}
   \end{figure}

        Although SCAN predicts the correct ZB ground state, its
        electronic PDOS spectra are very similar to those of PBESol
        (see Fig.~\ref{fig4:dos_scan}). In fact, it has been
        already pointed out that meta-GGAs
        (including SCAN) do not really improve over GGA for electronic
        and magnetic properties (band gaps, magnetic moments)
        ~\cite{isaacs18,fu19,sokolov23}.

   	Within SCAN framework, MnC is also predicted as a
        ferromagnetic metal in rocksalt structure and a ferromagnetic
        half-metal in zincblende one. The whole distribution of the
        bonding, non-bonding and anti-bonding regions is almost the
        identical to PBESol. The only noticeable difference is that the
        Fermi level is slightly pushed upward at the vicinity of a
        non-vanishing DOS, placing the compound near a fully metallic
        state. Such sensitivity to the lattice constant (here SCAN:
        4.28\AA~ versus PBESol: 4.22\AA) has been mentioned
        previously, where a small increase drives zincblende MnC from
        half-metallic to metallic state~\cite{sasioglu05,pask03}.
        
   \subsection{Hybrid PBE0r and HSE06}
   \label{pbe0r-hse06_results}

\begin{figure}[htb]
     \includegraphics[width=0.85\columnwidth,height=12cm]{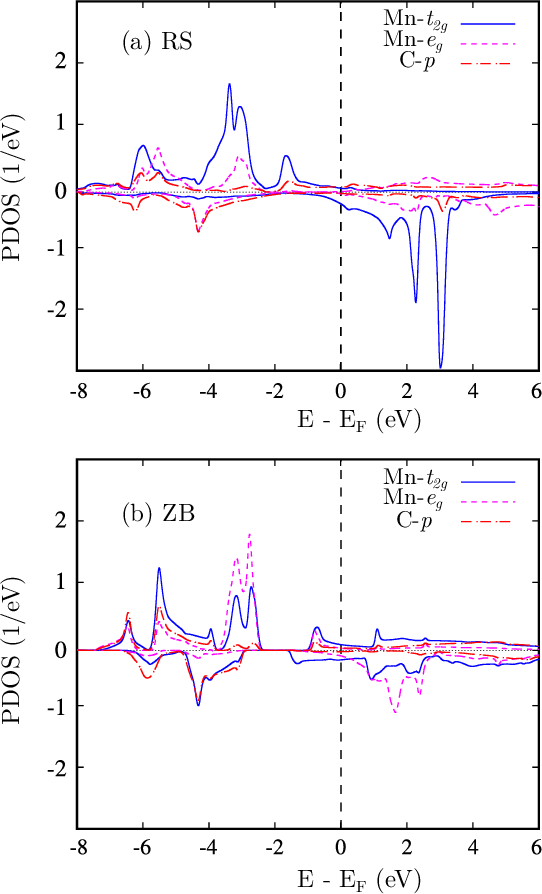}
     \caption{Atom- and spin resolved densities of states (PDOS) of $
       {\rm MnC} $ in RS-type (a) and ZB-type (b) structures, as
       arising from the PBE0r, with {\it bond} exchange (offsite=T),
       spin-polarized FM calculations.}
     \label{fig5:dos_pbe0r}
   \end{figure}

\begin{figure}[htb]
     \includegraphics[width=0.85\columnwidth,height=12cm]{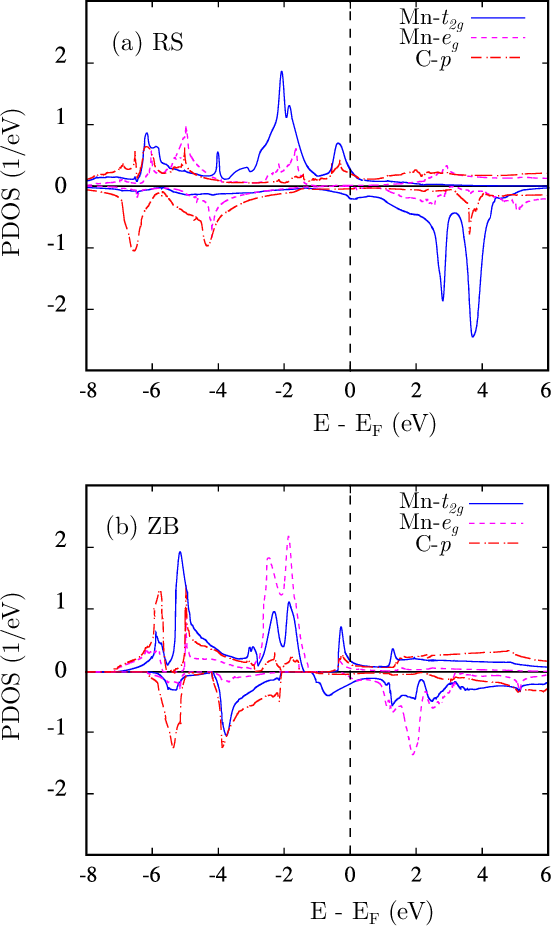}
     \caption{Atom- and spin resolved densities of states (PDOS) of $
       {\rm MnC} $ in RS-type (a) and ZB-type (b) structures, as
       arising from the HSE06 spin-polarized FM calculations.}
     \label{fig6:dos_hse06}
   \end{figure}

   	To assess the effect of a partial inclusion of the non-local
        HF exchange; we used PBE0r and HSE06 hybrid functionals, with
        two different fractions 10\% and 30\%. In
        each crystalline phase, both fractions yield typically
        similar trends in the electronic PDOS (bonding, non-bonding
        and anti-bonding regions). The only difference is the stronger
        exchange splitting with the larger fraction, producing larger
        magnetic moments. The electronic PDOS obtained from PBE0r and
        HSE06 are displayed in Fig.~\ref{fig5:dos_pbe0r} and
        Fig.~\ref{fig6:dos_hse06}, respectively. As a first results,
        the two functionals lead to a metallic MnC in both RS and ZB
        structures.

        In the rocksalt phase, the inclusion of exact exchange
        substantially enhances the spin splitting of the Mn-3{\it d}
        states, especially for the non-bonding Mn-{\it t$_{2g}$} ones.
        In PBE0r, the corresponding shift is more pronounced, so that
        the majority (minority) spin states are almost filled
        (empty). This drives the Mn-{\it t$_{2g}$} manifold
        (three-fold degenerate) to a quasi high-spin configuration,
        whereby the value of the magnetic moment $\simeq$ 3 $\mu_{\rm
          B}$ (see Tab.~\ref{table2}). In HSE06, a similar situation
        is found but with a slightly less important splitting.


   	In the zincblende structure, the same description holds for
        the two hybrid functionals. The Mn-{\it e$_g$} manifold
        (two-fold degenerate) experience the largest exchange
        splitting driving it toward a high-spin configuration,
        consistently with the computed magnetic moments. However, in
        contrast to the semilocal functionals, both hybrid PBE0r and
        HSE06 predict a fully metallic ZB-MnC at theoretical
        equilibrium. The disappearance of the half-metallicity is
        primarily due to lattice expansion, since both hybrid
        functionals yield larger equilibrium volumes (see
        Tab.~\ref{table2}); in complete agreement with GGA
        predictions.

        To confirm this result, additional PBE0r and HSE06
        calculations have been carried out; at smaller lattice
        constants, as obtained from PBESol and SCAN. The obtained
        results (not shown here), indicate that the majority-spin gap reappears
        and the half-metallicity is restored. Thus, all functionals
        agree that ZB-MnC is a half-metal at reduced volumes
        ($\leq$ 4.3 $\rm \AA$); and becomes metallic under
        expansion.  Consequently, the compound is expected to behave
        as a ferromagnetic metal at the experimental lattice constant
        ($\sim$ 4.4 $\rm \AA$).

   \subsection{Mean field DFT+$U$+$V$}
   \label{pbesol+u+v_results}
   \begin{table}[htb]
     \caption{The calculated values of the onsite $U$ and the
       intersite $V$ Hubbard parameters for Mn 3{\it d} states, in eV,
       at: DFT+$U$+$V$ and PBESol lattice constants; as well as
       the experimental one. The values are for ferromagnetic solution
       only; in the rocksalt (RS) and zincblende (ZB) structures.}
     \begin{ruledtabular}
       \begin{tabular}{c|cccc}
                          &       &    RS         &   ZB         \\
         \hline
          a $_{{\rm DFT}+U+V}$ &$ U $   &  5.60   &  5.50   \\
                                 &$ V $   &  0.57   &  0.70   \\
         \hline
          a $_{\rm PBESol}$         &$ U $   &  5.93   &  5.85   \\
                                 &$ V $   &  0.64   &  0.81   \\
          \hline
          a $_{\rm Exp.}$    &$ U $   &              &  5.53   \\
                                 &$ V $   &        &  0.76   \\  
       \end{tabular}
       \label{table3}
     \end{ruledtabular}
   \end{table}
   	The DFT+$U$+$V$ framework offers a complementary picture by
        explicitly introducing on-site $U$ and intersite $V$ Coulomb
        interactions. These parameters are computed self-consistently
        using the linear response DFPT method; and the obtained values
        at theoretical equilibrium are shown in Tab~\ref{table3}. For
        sake of completeness, we also provide the computed values at
        the PBESol equilibrium, and at the zincblende experimental
        lattice constant (4.43~\AA). While the on-site $U$ remains
        nearly constant (5.5–5.9 eV), the intersite term $V$ decreases
        with increasing lattice constant, reflecting the reduced
        overlap between neighboring Mn and C orbitals. These values
        are somewhat smaller than those typically found in
        transition-metal monoxides (For example, in MnO and NiO the
        $U$ parameter is larger than 6~eV
        \cite{tran06,seth17,mandal19}), confirming the weaker
        electronic correlations in MnC.

   \begin{figure}[htb]
     \includegraphics[width=0.85\columnwidth,height=12cm]{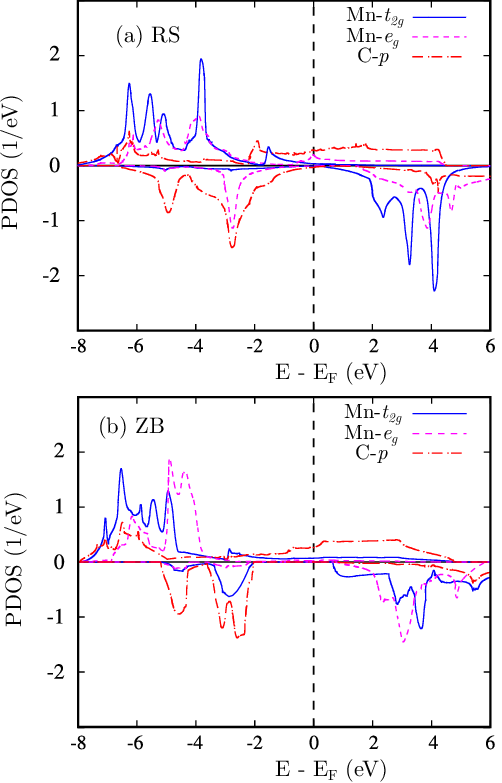}
     \caption{Atom- and spin resolved densities of states (PDOS) of $
       {\rm MnC} $ in RS-type (a) and ZB-type (b) structures, as
       arising from the DFT+$U$+$V$ spin-polarized FM calculations.}
     \label{fig7:dos_pbesol+u+v}
   \end{figure}

        The PDOS obtained from DFT+$U$+$V$ (shown in
        Fig.~\ref{fig7:dos_pbesol+u+v}) reveals several distinctive
        features. 
        In the rocksalt phase, the Mn-{\it t$_{2g}$}
        manifold (three-fold degenerate) is fully polarized i.e. in a
        complete high-spin configuration, giving rise to the largest
        computed magnetic moment value ($\rm M_{fu} \geqslant$ 3 $\mu_{\rm B}$).

        In the zincblende phase, the system is in a half-metallic
        state with a moderate minority-spin gap ($\sim$ 2.5 eV), in
        contrast to the predictions of hybrid PBE0r and HSE06. The
        total moment is 3 $\mu_{\rm B}$, where the high-spin 
        Mn-{\it e$_g$} manifold contribute with 2 $\mu_{\rm B}$ and
        the remaining 1 $\mu_{\rm B}$ comes mainly from Mn-{\it
          t$_{2g}$} states.

        Interestingly, while semilocal functionals and DFT+$U$+$V$
        agree on the half-metallicity, the band gap resides in
        opposite spin channels. Curiously, in DFT+$U$+$V$ this
        behavior occurs at a lattice constant (4.55~\AA) much larger
        than the limit ($\leq$ 4.30~\AA) predicted by semi-local
        functionals. Additional DFT+$U$+$V$ calculations (using the
        corresponding calculated parameters, see Tab.~\ref{table3})
        show that the half-metallicity persists at the experimental
        lattice constant (4.43~\AA), and vanishes only under strong
        compression at the PBESol volume. This opposite volume trend
        highlights fundamental differences on how semilocal and hybrid
        functionals in one side, and Hubbard-corrected ones in the other,
        treat the MnC half-metallic behavior.

        In this context, a similar observation has been made for
        several compounds crystallizing in the 
        zincblende structure. For many pnictides, such as
        Mn(Cr)As(Sb), GGA calculations predicted a half-metallic
        behavior with a minority-spin gap, disappearing under
        compression~\cite{pask03}. In the same work, MnC was the
        only half-metal with majority-spin gap.

        \subsection{Discussion}
         \label{discussion}

        Across all theoretical frameworks, the Mn(3{\it d})--C(2{\it
          p}) hybridization defines the essential features about the
        bonding properties and the electronic structure of MnC. Our
        comprehensive calculations reveal some disagreements
        on the electronic ground state of ZB-MnC. This is not a
        failure of the methods but a reflection of the compound's
        sensitive position at the boundary between a half-metal and a
        correlated metal. The predictions fall into two distinct
        categories:

        1. Hybrid and semilocal functionals agree that
        half-metallicity appears at small volume and vanishes upon a
        slight expansion ($\sim$4.30~\AA). At the experimental lattice
        constant ($\sim$4.43~\AA), accurately provided by PBE0r, MnC
        is consistently predicted as a metal. The non-local
        exact exchange enhances exchange splitting and magnetic moments.

        2. On the other side, DFT+$U$+$V$ favors half-metallicity at
        large volume, but with a different nature. It predicts a
        robust half-metallic state, with a moderate minority-spin gap,
        persisting at the experimental volume and disappears only
        under strong compression. 

        This divergence creates a clear theoretical controversy, where
        the more structurally accurate hybrid functionals suggest a
        metal; and methods that often better describe localized states
        suggest half-metallicity. Such controversy could be resolved
        by experiment, with specific techniques:

        $\bullet$ A direct measurement of the spin polarization at the
        Fermi level, via spin-resolved photoemission spectroscopy
        (SPES) or angle-resolved photoemission (ARPES) to confirm
        Metallic vs. Half-Metallic behavior.

        $\bullet$ High-pressure electrical transport and magnetization
        measurements could track the volume dependence of the
        conductivity and magnetic moment.

        Therefore, the true electronic ground state of MnC remains an
        exciting open question, whose resolution will have significant
        implications for its practical applications.
 
   \section{Conclusion}
   \label{conclusion}

   	A comprehensive first-principles investigation of MnC
        properties was carried out, using the state-of-the-art of DFT
        based methods: PBESol, SCAN, hybrid functionals (PBE0r and
        HSE06), and the mean field DFT+$U$+$V$. The main goal was to
        identify the origin of the experimentally observed zincblende
        structure, and to examine the role of bond exchange in its
        structual stability as well as electronic behavior.

        Our results demonstrate that the stabilization of this
        experimental structure is uniquely achieved by the SCAN
        meta-GGA and the PBE0r hybrid functional. This success is
        attributed to the improved (implicit or explicit) treatment of
        {\it bond-directional exchange} interactions in covalent
        systems. The decisive role of this exchange term is
        quantitatively proven by its explicit inclusion in PBE0r,
        which selectively stabilizes the tetrahedral ZB phase by
        favoring bonding over antibonding states. This shows that an
        accurate orbital-selective treatment of exchange is crucial
        for describing the structural chemistry of MnC.

        Beyond crystal structure, our investigation reveals a
        fundamental issue in predicting the electronic ground
        state. While semilocal functionals (PBESol, SCAN) and
        DFT+$U$+$V$ predict integer momnets and half-metallicity,
        albeit in opposite spin channel, hybrid functionals (HSE06,
        PBE0r) yield larger magnetic moments and metallic state. The
        electronic character is found to be strongly volume-dependent:
        semi-local and hybrid functionals point towards a disappearance
        of the half-metallicity under expansion, but the opposite is
        observed with the DFT+$U$+$V$ approach. This discrepancy
        highlights the sensitive position of MnC at the boundary
        between a correlated metal and a half-metal. It establishes
        the material's electronic state as an open question for both
        theory and experiment.

        Therefore, we conclude that the true ground state of ZB-MnC is
        best described by a framework that prioritizes bond exchange,
        such as SCAN or PBE0r. These methods consistently lead
        to a ferromagnetic metal at ambient synthesis
        conditions, with a potential transition to a half-metallic
        state under compression. Taken together, these findings
        establish MnC as a model system for studying the interplay
        between covalent bonding and exchange interactions in late
        3{\it d} transition-metal compounds. Finally, this work not
        only clarifies the properties of a newly synthesized material
        but also proposes a clear roadmap for its experimental
        validation.

\end{document}